# A Compact Virtual-Source Model for Carbon Nanotube Field-Effect Transistors in the Sub-10-nm Regime—Part I: Intrinsic Elements

Chi-Shuen Lee, Eric Pop, *Senior Member, IEEE*, Aaron D. Franklin, *Senior Member*, *IEEE*, Wilfried Haensch, *Fellow*, *IEEE*, and H.-S. Philip Wong, *Fellow, IEEE*

*Abstract*—We presents a data-calibrated compact model of carbon nanotube (CNT) field-effect transistors (CNFETs) based on the virtual-source (VS) approach, describing the intrinsic current-voltage and charge-voltage characteristics. The features of the model include: (i) carrier VS velocity extracted from experimental devices with gate lengths down to 15 nm; (ii) carrier effective mobility and velocity depending on the CNT diameter; (iii) short channel effect such as inverse subthreshold slope degradation and drain-induced barrier lowering depending on the device dimensions; (iv) small-signal capacitances including the CNT quantum capacitance effect to account for the decreasing gate capacitance at high gate bias. The CNFET model captures dimensional scaling effects and is suitable for technology benchmarking and performance projection at the sub-10-nm technology nodes.

## I. INTRODUCTION

CARBON nanotube field-effect transistors (CNFETs) based on single-walled semiconducting CNTs have been among the foremost candidates to complement Si and extend CMOS technology scaling in the sub-10-nm technology nodes [1-3]. One of the dominant factors impeding further scaling of Si metal-oxide-semiconductor field-effect transistors (MOSFETs) is the short-channel effect (SCE), which causes FETs at short gate lengths to be difficult to turn off, consequently consuming too much power [4]. To keep Si MOSFETs viable, a tremendous amount of effort has been put into transitioning from two-dimensional (2D) planar device structures to three-dimensional (3D) channel geometry with fin structures [5]. However, further scaling the gate length ($L_g$) of Si-MOSFETs requires an ultra-thin channel body [4], resulting in low drive current due to mobility degradation (caused by the body thickness fluctuation [6]) and low density of states (DOS) [7].

By contrast to bulk 3D materials, a single-walled CNT is essentially a single sheet of graphene rolled into a seamless cylinder with a 1-2 nm diameter. Because of the atomically thin body, the gate control of CNFETs is superior and the SCE can be overcome even for $L_g$ < 10 nm [1-2,8]. Furthermore, CNTs show promise for energy-efficient computation because of their high carrier velocity and near-ballistic carrier transport property [9-10]. A great deal of effort has been made to demonstrate the potential of CNFETs as the future transistor technology. Recent progress in CNFET technology include a 9-nm $L_g$ CNFET [2], realization of a gate-all-around (GAA) device [11], complementary n- and p-type CNFETs [12], operation at low (0.4 V) voltages [13], and the demonstration of a simple CNT computer [14]. Despite their great potential, CNFETs suffer from imperfections such as difficulty in obtaining purely semiconducting CNTs [15], hysteresis of the current-voltage (I-V) characteristics [16], mis-positioning and diameter variations [17]. Several techniques have been reported to overcome these imperfections from the fabrication process level up to the system architecture level [18], and more improvement is needed to realize CNFET-based electronics.

For all emerging technologies, early assessment based on both experimental observation and theoretical study is of great value as it facilitates identification of the most promising options and allows resources to be focused on them. Non-equilibrium Green's function (NEGF) formalism [19], recognized as a physically rigorous approach, has been extensively employed to simulate quantum transport in CNFETs and assess their performance [8,20]. However, NEGF is too computationally expensive for performance assessment at the application level. Compact modeling based on the Landauer formula for ballistic transport in CNTs is another efficient approach for performance assessment of CNFETs [21-23]. While some of these compact models have been validated by numerical simulation or experimental data, the effects of dimensional scaling, series resistance ($R_s$), and tunneling leakage current have not been well captured. Attempts were made to address this issue by lumping the scaling and parasitic effects into constant input parameters (e.g. constant $R_s$, mobility, and subthreshold slope) independent of the device design [21].

Manuscript received Mar. xx, 2015. This work was supported in part through the NCN-NEEDS program, which is funded by the National Science Foundation, contract 1227020-EEC, and by the Semiconductor Research Corporation, and through Systems on Nanoscale Information fabriCs (SONIC), one of the six SRC STARnet Centers, sponsored by MARCO and DARPA, as well as the member companies of the Initiative for Nanoscale Materials and Processes (INMP) at Stanford University.

C. -S. Lee, E. Pop, and H.-S. P. Wong are with the Department of Electrical Engineering, Stanford University, Stanford, CA 94305 USA (e-mail: chishuen@stanford.edu; hspwong@stanford.edu).
A. D. Franklin is with the Department of Electrical and Computer Engineering, Duke University, Durham, NC 27708 (e-mail: aaron.franklin@duke.edu).
W. Haensch is with IBM T. J. Watson Research Center, Yorktown Heights, NY 10598 USA (e-mail: whaensch@us.ibm.com).

As a result, the dimensional scaling effect and variations cannot be studied.

In this paper, we describe a data-calibrated compact CNFET model based on the virtual source (VS) approach [24], which has been implemented in Verilog-A [25] and available online [26]. The main motivation of developing the VS-CNFET are two-fold: (i) to identify the required improvement in device and materials to achieve performance advantage over similarly scaled FETs, and (ii) to enable performance assessment of CNFET systems at the application level, including device non-idealities and variations. In the model, the VS parameters (e.g. carrier mobility, velocity, and gate capacitance) are connected to the CNFET dimensions and CNT diameter in order to capture the scaling effect. A similar concept has been reported in [27] with preliminary results that did not include several important effects: small-signal capacitances were not properly modeled; CNT quantum capacitance was not considered; the internal VS parameters were independent of CNT diameter; iterations and numerical integral were needed. These deficiencies are addressed in this paper.

Several premises are relied upon in this work: (i) we focus purely on MOSFET-like CNFETs with Ohmic metal-CNT contacts, because they provide better performance and could be realized by heavily doping the source/drain (S/D) extensions [9,28]. Previous efforts on modeling Schottky-barrier CNFETs can be found in [29]; (ii) n-type CNFETs are discussed throughout the paper. Although CNFETs in ambient air are usually p-type based on the preferred injection of holes at the contacts, n-type CNFETs have been achieved by contact or interface engineering [30-31], and from a physical and mathematical point of view the operation of n-type and p-type CNFETs is symmetric due to the symmetry of conduction and valence bands; (iii) only the first sub-band in CNTs is considered because most digital applications call for a low power supply voltage, but higher sub-bands can be easily included with proper modification of the charge model.

This paper is organized as follows: analytical expressions that connect the VS parameters to the CNFET design as well as model calibration are described in Section II; the charge model used to derive the small-signal capacitances is introduced in Section III; in Section IV, the impact of CNT diameter on the intrinsic CNFET performance is presented; finally in Section V, issues pertaining to the VS parameter extraction from CNFETs are discussed. Due to the limited space, the complete derivation of all the equations is detailed in [26]; here we only discuss the physics and key results. Models for the extrinsic elements such as contact resistance and tunneling leakage current will be introduced in Part II of this two-part paper [32].

II. VIRTUAL SOURCE MODEL FOR CNFETs

The VS model is a semi-empirical model with only a few physical parameters, originally developed for short-channel Si MOSFETs that have a gate-controlled source-injection barrier. Recently an enhanced VS emission-diffusion model applicable to both long-channel FETs (in drift-diffusive carrier transport regime) and short-channel FETs (in quasi-ballistic regime) has been reported [33]. Here the VS-CNFET model is based on the VS model described in [24,34] because we focus on the short-channel FETs (e.g. $L_g < 30$ nm) where the carrier transport is assumed to be quasi-ballistic. Based on the VS approach, the drain current ($I_d$) of a MOSFET is the product of the mobile charge density and the carrier velocity at the VS, defined as the top of the energy barrier near the source in the on-state, where the lateral electric field is small and the potential is mostly controlled by the gate voltage [24]. There are ten VS parameters: gate length ($L_g$); gate capacitance in strong inversion region ($C_{inv}$); low-field effective mobility ($\mu$); threshold voltage ($V_t$); inverse subthreshold slope (SS) factor ($n_{ss}$); drain-induced barrier lowering (DIBL) coefficient ($\delta$); series resistance ($R_s$); VS carrier velocity ($v_{xo}$); and fitting parameters $\alpha$ and $\beta$ used to smooth the transitions between weak and strong inversion, and between non-saturation and saturation regions, respectively. As conceived originally, the VS model was not meant to be predictive because the VS parameters need to be extracted from current-voltage (I-V) and capacitance-voltage (C-V) measurements; the VS-CNFET model instead associates the VS parameters to the device dimension and CNT diameter so that the CNFET design is connected to the device-level characteristics. Furthermore, by calibrating the VS-CNFET model to experimental data and rigorous numerical simulations, it becomes possible to make predictive estimates of device behavior as the dimension scales down. A representative GAA-CNFET device structure used in the VS-CNFET model is illustrated in Fig. 1 with the critical dimensions labeled. In this section, analytical models to bridge the VS parameters to the device dimension and CNT diameter are introduced.

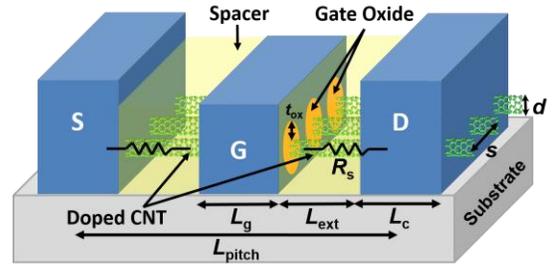

Fig. 1. A representative gate-all-around CNFET device structure used in the VS-CNFET model with the critical dimensions labeled.

CNT diameter ($d$) is a crucial physical parameter because it determines the CNT band structure and the band gap ($E_g$). In this work, CNT $E_g = 2E_p a_{cc}/d$ is derived from the Hückel tight-binding model [35], where $E_p = 3$ eV is the tight-binding parameter, and $a_{cc} = 0.142$ nm is the carbon-carbon distance in CNTs, indicating $E_g \approx 0.85/d$ eV with $d$ in nm. Corrections to the model of $E_g$ could be made due to band gap renormalization induced by many-body interaction [36] or substrate-induced polarization effects [37], but they do not alter the essence of the VS model presented here. As will be shown later in this section, $C_{inv}$, $\mu$, $V_t$, $n_{ss}$, $\delta$, and $v_{xo}$ are all diameter-dependent in the VS-CNFET model.

A. *Inversion Gate Capacitance ($C_{inv}$)*

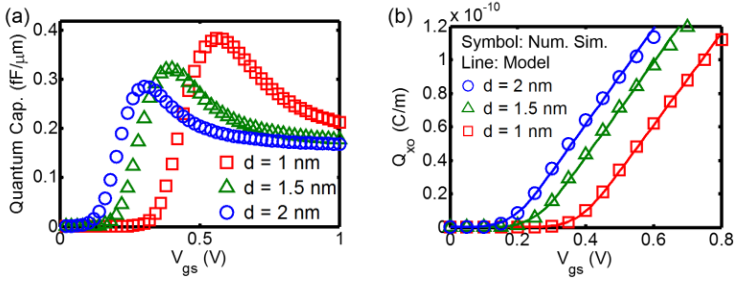
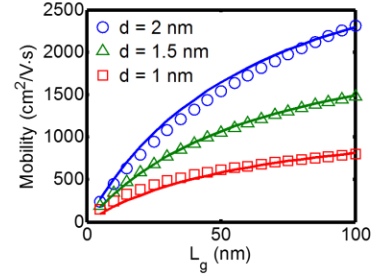

Fig. 2. (a) Numerical simulated CNT quantum capacitance $C_q$ vs. $V_{gs}$ for different diameters $d$. The peak $C_q$ increases as $d$ decreases. (b) Comparison of the VS carrier density $Q_{xo}$ between the numerical simulation [41] and the model given (see II.A and [26]). Equivalent oxide thickness (EOT) = 0.7 nm.

Fig. 3. Low-field mobility vs. $L_g$ for different diameters. The symbols are the peak mobility calculated numerically by (2) and the lines represent the model given by (4). The mobility decreases towards smaller $L_g$, as the conductance becomes constant with quasi-ballistic transport, see (2a).

In a MOSFET, the mobile charge density in strong inversion at the VS, where the gradual channel approximation applies [38], can be approximated as $Q_{xo} \approx -C_{inv} \cdot (V_{gs} - V_t)$, where $C_{inv} = C_{ox} \cdot C_s / (C_{ox} + C_s)$, $C_{ox}$ is the gate oxide, $C_s$ is the semiconductor capacitance defined as $-dQ_{xo}/d\psi_s$, and $\psi_s$ is the surface potential [39]. For a GAA structure, $C_{ox}$ is:

$$C_{ox} = \frac{2\pi k_{ox} \varepsilon_0}{\ln\left[(2t_{ox} + d)/d\right]} \quad (1)$$

where $\varepsilon_0$ is the permittivity in vacuum, $t_{ox}$ and $k_{ox}$ are the thickness and the relative dielectric constant of the gate oxide, respectively. In planar bulk semiconductor materials, the DOS is usually so large that $C_s \gg C_{ox}$ and $C_{inv} \approx C_{ox}$; however, for CNTs, the CNT quantum capacitance ($C_q$) needs to be considered because $C_q$ is comparable to $C_{ox}$ due to the relatively low DOS. In the VS-CNFET model, $C_{inv} = C_{ox} \cdot C_{qe}/(C_{ox} + C_{qe})$, where $C_{qe}$ is an empirical parameter representing an effective CNT quantum capacitance. Strictly speaking, $C_q$ is bias dependent [40] as shown in Fig. 2a. However, the numerical simulation in Fig. 2b shows that the linear relation between $Q_{xo}$ and $V_{gs} - V_t$ in the inversion region is still retained over a reasonable range of $V_{gs}$ and $Q_{xo}$, implying the viability of having a constant $C_{qe}$ to account for the effect of quantum capacitance. The numerical simulator used to validate the VS-CNFET model throughout this paper is provided by [41], which simulates a GAA-CNFET with heavily doped S/D regions, and the carrier transport is simulated based on the NEGF formalism. It has been shown in [40] that the maximum CNT $C_q$ is approximately proportional to $E_g^{1/2}$; here $C_{qe}$ is empirically modeled as $C_{qe} = 0.64 E_g^{1/2} + 0.1$ (fF/µm), where $E_g$ is in the unit of eV, and the coefficients are extracted by fitting the modeled $Q_{xo}$ to the numerical simulation in Fig. 2b. The modeled $Q_{xo}$ is calculated by substituting $C_{inv}$ into the equation in [26, Eq. (1.3)].

*B. Carrier Mobility ($\mu$)*

As $L_g$ scales down to nanoscale, the carrier transport approaches the ballistic limit and carrier scattering in the channel becomes less significant. In this paper, the mobility is the so-called apparent mobility [38], a concept that connects the ballistic and diffusive regimes. The apparent mobility could also be understood as another way to express the mean free path (MFP). As device dimensions become smaller than the MFP, the carriers travel across the channel nearly without scattering and scatter only at the source and drain. In this context, the MFP becomes the channel length.

In the VS-CNFET model, $\mu = GL_g/(qn_s)$ is derived from the one dimensional (1D) quantum transport theory at low fields, written here for the lowest sub-band [19,42]:

$$G = \frac{4q^2}{h} \int_{E_c}^{\infty} \frac{\lambda_i(E)}{L_g + \lambda_i(E)} \left[-\frac{\partial f(E, E_F)}{\partial E}\right] dE \quad (2a)$$

$$n_s = \int_{E_c}^{\infty} g(E) f(E, E_F) dE \quad (2b)$$

where $G$ is the CNT conductance, $n_s$ is the carrier density, $q$ is the elementary charge, $h$ is Planck's constant, $E_c$ is the conduction band edge, $E$ is the energy of free electrons, $E_F$ is the Fermi level, $f$ is the Fermi-Dirac distribution function, $g(E)$ is the CNT DOS, and $\lambda_i$ is the MFP in CNTs representing the aggregate effect of optical phonon (OP) and acoustic phonon (AP) scattering [42]:

$$\frac{1}{\lambda_i} = \frac{1}{\lambda_{AP}(E,T)} + \frac{1 - f(E + \hbar\omega_{OP})}{\lambda_{OP,abs}(E,T)} + \frac{1 - f(E - \hbar\omega_{OP})}{\lambda_{OP,ems}(E,T)} \quad (3)$$

where $T$ is the temperature, $\hbar\omega_{OP} \approx 0.18$ eV is the OP energy, $\lambda_{AP}$, $\lambda_{OP,abs}$ and $\lambda_{OP,ems}$ are MFPs for AP scattering, OP absorption and emission, respectively. The expression for $\lambda_i$, its experimental validation, and treatment across multiple sub-bands have been detailed in [42] (only the lowest sub-band is considered here). However, due to the complex expression for $\lambda_i$, Eq. (2a) cannot be integrated analytically; to avoid the use of a numerical integral in the compact model, an empirical expression is used to model $\mu$:

$$\mu = \mu_0 \frac{L_g}{\lambda_\mu + L_g} (d)^{c_\mu} \quad (4)$$

where $d$ is in the unit of nm, $\mu_0 = 1350$ cm$^2$/V·s, $\lambda_\mu = 66.2$ nm, and $c_\mu = 1.5$ are empirical parameters extracted by fitting (4) to the peak mobility (note that $\mu$ in (2a) depends on $E_F$) calculated by (2) and (3) as shown in Fig. 3. It should be noted that for device configurations similar to Fig. 1, the source and drain are in fact separated by $L_g + 2L_{ext}$ rather than $L_g$. However, since the extensions are not gated and have higher doping densities than the region under the gate (thus different MFPs), we treat the extensions in [32] as extrinsic elements and confine the scope of intrinsic elements (described by the VS model) to the region under the gate, leading to a hierarchical model. In experimental measurements, however, it is not easy to separate the region under the gate from the extensions and the contacts;

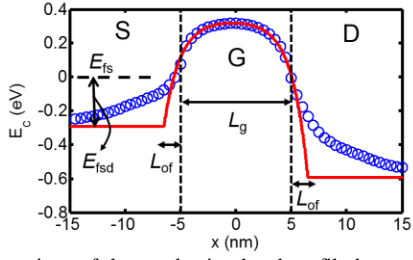

Fig. 4. Comparison of the conduction band profile between the numerical simulation [41] (symbols) and the model (line) given by (5).

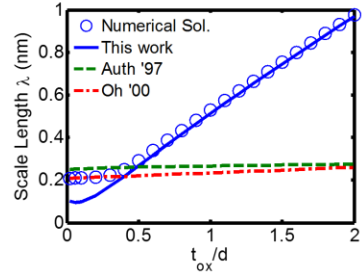

Fig. 5. Comparison of the scale length $\lambda$ between the numerical solution to (6) and the model given by (7) for $d = 1$ nm, $k_{ox} = 16$, and $k_{cnt} = 1$. Good agreement is observed when $t_{ox} > d/2$, while the approximation given in [45] works better for $t_{ox} < d/2$.

hence any extraction of mobility for a short-channel CNFET from I-V measurements is actually a reflection of the commingled behaviors of contact injection and carrier transport in the extensions and the channel. Therefore, the use of apparent mobility [38] in the VS model can be viewed as a convenience for describing the experimental I-V curves in a hierarchical model. We note the apparent mobility approaches zero as the channel length (which limits the MFP) approaches zero, consistent with the ballistic limit.

*C. SCE Parameters (SS, DIBL, $V_t$ roll-off)*

The SCE is essentially the phenomenon of decreasing $V_t$ and increasing SS and DIBL as $L_g$ scales down. In this paper, the SCE parameters are derived from a GAA cylindrical structure based on the scale length theory [43]. The first step is to model the conduction band ($E_c$) profile along the channel. In the subthreshold region where the mobile charge in the channel is negligible, the $E_c$ profile can be obtained by solving the Laplace equation, and the resulting $E_c$ can be expressed as:

$$E_c(x) = a_1 e^{-x/\lambda} + a_2 e^{x/\lambda} - V_{gs} + E_g/2 \qquad (5)$$

where $x$ is the direction along the channel, $\lambda$ is the electrostatic scale length (also known as the screening length), and $a_1$ and $a_2$ are coefficients determined by the boundary conditions: $E_c(-L_{of}-L_g/2) = -E_{fsd}$ and $E_c(L_{of}+L_g/2) = -E_{fsd}-V_{ds}$, where $L_{of}$ is an empirical parameter functioning like an extension of the $L_g$ that captures the finite Debye length at the gate-to-source/drain junctions, and $E_{fsd}$ is the energy difference from the Fermi level to the $E_c$ at the S/D extensions (see Fig. 4). All energies are referenced to the Fermi level at the source (i.e. $E_{fs} = 0$).

In a GAA cylindrical structure, $\lambda$ is a solution to the Laplace equation in cylindrical coordinates satisfying the boundary condition at the CNT/oxide interface:

$$\frac{Y_1(\zeta)}{J_1(\zeta)} = \gamma \frac{Y_0(\zeta)}{J_0(\zeta)} + (1-\gamma)\frac{Y_0(\zeta + t_{ox}/\lambda)}{J_0(\zeta + t_{ox}/\lambda)} \qquad (6)$$

where $J_m$ and $Y_m$ are Bessel functions of the first kind and second kind of order $m$, $\gamma \equiv k_{cnt}/k_{ox}$, $k_{cnt}$ is the relative dielectric constant of the CNT, and $\zeta \equiv d/(2\lambda)$. Eq. (6) is a transcendental equation which has no closed form solution for $\lambda$. Analytical approximations of $\lambda$ in GAA-MOSFETs have been derived in [44] by assuming that the $E_c$ profile is parabolic in the transverse direction; however for CNFETs, $d$ is often smaller than $t_{ox}$, so the approximation made in [44] fails. When $t_{ox} > d/2$, we show that $\lambda$ can be approximated as:

$$\lambda = \frac{d+2t_{ox}}{2z_0}\left[1+b(\gamma-1)\right]$$
$$b = 0.41\left(\zeta_0/2 - \zeta_0^3/16\right)(\pi\zeta_0/2) \qquad (7)$$
$$\zeta_0 = z_0 d/(d+2t_{ox})$$

where $z_0 \approx 2.405$ is the first zero of $J_0$. Derivation of (7) is detailed in [26, Eq. (15)-(19)]. Eq. (7) is compared with the numerical solution to (6) in Fig. 5, showing good agreement when $t_{ox} > d/2$. When $t_{ox} \gg d$, Eq. (7) can be simplified to $\lambda \approx (d+2t_{ox})/z_0$; on the other hand, when $t_{ox} \ll d$, it has been shown in [45] that $\lambda \approx (d+2\gamma \cdot t_{ox})/z_0$. In both extreme cases, $\lambda$ increases linearly with $d$ and $t_{ox}$. In this paper, $k_{cnt} = 1$ is used, assuming it is air inside the CNT [46]. However, different values of $k_{cnt}$ from 5-10 for semiconducting CNTs have been reported both theoretically [47] and experimentally [48]. Nonetheless, we can show that (7) holds for a wide range of $k_{cnt}$ (from 1~20).

By substituting (7) into (5), $E_c$ profile is calculated and compared to the numerical simulation [41] in Fig. 4, showing good agreement in the gate region. Although the potential "tails" extending into the S/D extensions are not captured by (5), this will not affect the calculation of SCE parameters since only the top of the $E_c$ ($E_{cmax}$) matters. Modeling of the tails will be discussed in [32] when calculating the tunneling currents. Once the $E_c$ profile is known, the SCE parameters can be derived as:

$$n_{ss} = -\partial E_{cmax}/\partial V_{gs}\big|_{V_{ds}=0} = (1-e^{-\eta})^{-1} \qquad (8.1)$$

$$\delta = -\partial E_{cmax}/\partial V_{ds}\big|_{V_{ds}=0} = e^{-\eta} \qquad (8.2)$$

$$-\Delta V_t = E_g/2 - E_{cmax}\big|_{V_{ds}=0} = (2E_{fsd}+E_g)e^{-\eta} \qquad (8.3)$$

where $\eta \equiv (L_g+2L_{of})/2\lambda$, and $E_{cmax}$ is calculated by substituting $x = -\lambda/2 \cdot \ln(a_2/a_1)$ into (5). Eq. (8) is compared to the numerical simulation in Fig. 6. Empirically, $L_{of} \approx t_{ox}/3$ is found to achieve the best fitting results. A physical interpretation of the relation between $L_{of}$ and $t_{ox}$ is that when $t_{ox}$ becomes larger, the fringe field from the gate to the S/D extensions will extend, making $L_{of}$ longer. Nevertheless, generally $L_{of}$ should be viewed as a fitting parameter. Note that (8) is a direct result of solving Poisson's equation without considering non-idealities such as oxide-CNT interface states. Therefore, SS $\approx$ 60 mV/dec and DIBL = 0 for long-channel devices. More discussion on the oxide-CNT interface is included in [32]. Although (8) are derived from a GAA structure, other device structures such as top gate and bottom gate should follow the same trend as long as a proper model for $\lambda$ is used.

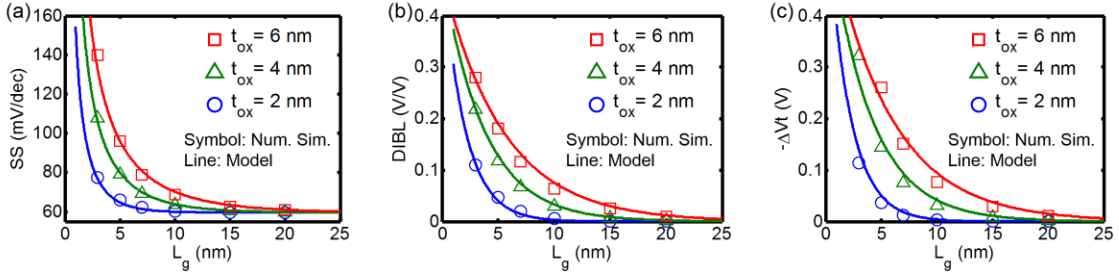

Fig. 6. Comparison of (a) SS, (b) DIBL, and (c) $V_t$ roll-off between the numerical simulation [41] and the model given by (8) for different gate oxide thickness. Tunneling currents are not excluded.

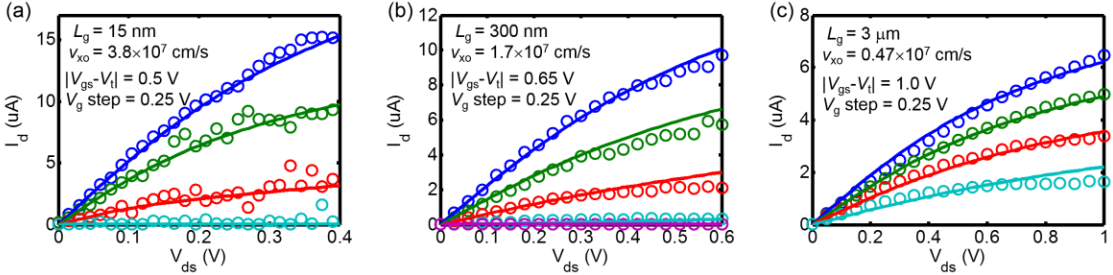

Fig. 7. Extraction of VS carrier velocity. The symbols are experimental data from [51]. (a) $v_{xo} = 3.8\times 10^7$ cm/s for $L_g = 15$ nm. (b) $v_{xo} = 1.7\times 10^7$ cm/s for $L_g = 300$ nm. (c) $v_{xo} = 0.47\times 10^7$ cm/s for $L_g = 3$ μm. Note that the polarity of $V_{gs}$ and $V_{ds}$ are flipped compared to the original data to become n-type FETs.

### D. Virtual Source Carrier Velocity ($v_{xo}$)

The VS carrier velocity ($v_{xo}$), also known as the injection velocity, is one of the key metrics for the transistor technology [49]. $v_{xo}$ can be associated with $L_g$ through the theory of back scattering of carriers in the channel [50]:

$$v_{xo} = \frac{\lambda_v}{\lambda_v + 2l} v_B \qquad (9)$$

where $v_B$ is the carrier velocity in the ballistic limit, $\lambda_v$ is the carrier MFP, and $l$ is the critical length defined as the distance over which the electric potential drops by $k_B T/q$ from the top of the energy barrier in the channel, where $k_B$ is the Boltzmann's constant. Strictly speaking, $l$ is proportional to $L_g$ and dependent on $V_{ds}$, as described in [33]. However, since using a bias-independent $v_{xo}$ can fit the experimental $I_d$-$V_{ds}$ data fairly well for different $L_g$'s (as will be seen shortly) and only a small range of $L_g$ is of our interest (e.g. 5 nm < $L_g$ < 30 nm), here $l \approx L_g$ is assumed for the sake of simplicity and $\lambda_v$ is thus empirical. To extract $v_B$ and $\lambda_v$, the VS model [24] is fitted to the $I_d$-$V_{ds}$ data from [51], where three CNFETs on the same substrate with identical structures but different gate lengths were measured.

The extraction flow of $v_{xo}$ involves: (a) $d = 1.2$ nm, $L_g = 15$ nm/300 nm/3 μm, $R_s = 5.5$ kΩ, and SS = 135 mV/dec according to the reported experimental data in [51]; (b) estimating, due to lack of C-V data, $C_{ox} = 0.156$ fF/μm by simulating a metallic cylinder placed on a 10-nm thick HfO$_2$ with a back gate using TCAD Sentaurus [52]; (c) $\mu = 255/10^3/2.1\times 10^3$ cm$^2$/V·s for $L_g = 15$ nm/300 nm/3 μm respectively, estimated by (2); (d) $\alpha = 3.5$ and $\beta = 1.8$ as suggested in [24]; (e) DIBL and $V_t$ are treated as free parameters because the two parameters are susceptible to the oxide-CNT and air-CNT interface properties and may suffer from different degrees of the hysteresis effect [16]. Fortunately, the extracted $v_{xo}$ is not sensitive to the choice for DIBL and $V_t$. Finally, $v_{xo}$ is treated as a free parameter to achieve the best fitting result as shown in Fig. 7. If uncertainty exists in the exact value of $d$ due to the measurement, the values of $C_{ox}$ and $\mu$ would be adjusted accordingly and the extracted $v_{xo}$ could be slightly different, but the change will be minor and the scaling trend will remain the same. By fitting (9) to the extracted $v_{xo}$'s, $\lambda_v = 440$ nm and $v_B = 4.1\times 10^7$ cm/s are extracted. $v_{xo}$ for other materials have been extracted from devices at various $L_g$'s, including $1.35\times 10^7$ cm/s for 32-nm $L_g$ Si MOSFET [24] and $3.2\times 10^7$ cm/s for 30-nm $L_g$ III-V HFET [53].

To model the dependence of $v_{xo}$ on CNT diameter, we refer to the carrier transport theory in MOSFETs [54]: the maximum value of $v_{xo}$ is approximately the equilibrium uni-directional thermal velocity $v_{Ti}$. For the non-degenerate case, $v_{Ti} = 2k_B T/(\pi m^*)$, where $m^* = h^2/(9\pi^2 a_{cc}^2 E_p d)$ is the effective mass in CNTs [40]. Therefore we can express $v_B = v_{B0}(d/d_0)^{1/2}$, where $v_{B0} = 4.1\times 10^7$ cm/s and $d_0 = 1.2$ nm are extracted from [2] set as reference points. To examine the validity of the linear relation between $v_B$ and $d^{1/2}$, the 1D Landauer formula [19] is used to calculate the theoretical ballistic velocity $v_{Bth}$:

$$\begin{aligned}
I_{dB} &= \frac{4q}{h}\int [f_S(E) - f_D(E)]\,dE \\
&= \frac{4q}{h} k_B T \ln\left[\frac{1+\exp\left(\dfrac{\psi_s - E_g/2q}{k_B T/q}\right)}{1+\exp\left(\dfrac{\psi_s - E_g/2q - V_{ds}}{k_B T/q}\right)}\right]
\end{aligned} \qquad (10)$$

where $I_{dB}$ is the drain current in the ballistic limit calculated by the 1D Landauer formula, and $v_{Bth} = I_{dB}/n_s$, where $n_s$ is calculated by (2b). Fig. 8 shows $v_{Bth}$ vs. $d^{1/2}$ for different carrier densities, indicating that the linear relation between $v_{Bth}$ and $d^{1/2}$ holds for a wide range of $d$ and $n_s$.

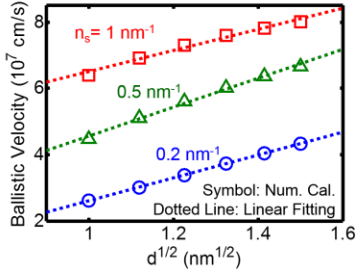
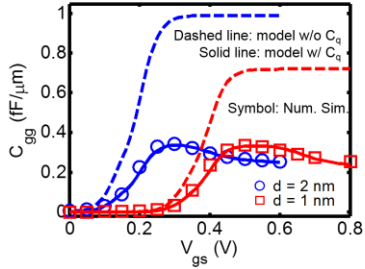
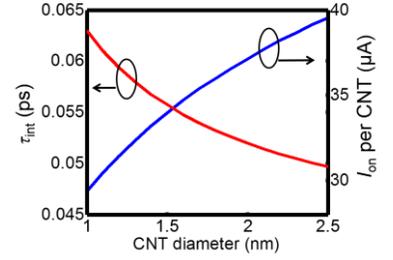

Fig. 8. Theoretical carrier velocity in the ballistic limit (symbols) vs. the square-root of CNT diameter (dotted lines) for different carrier densities ($n_s$). The symbols are calculated by (10) numerically.

Fig. 9. Comparison of small-signal gate capacitances $C_{gg}$ between the numerical simulation [41] and the model given by (11) at $V_{ds} = 0$. The dashed lines represent the case where CNT quantum capacitance is not considered.

Fig. 10. Intrinsic on-state current $I_{on}$ and gate delay $\tau_{int}$ vs. CNT diameter at $L_g = 8$ nm and $V_{dd} = 0.71$ V. A 2-nm diameter CNT has 27% higher $I_{on}$ and 21% lower $\tau_{int}$ than a 1-nm diameter CNT due to higher mobility, carrier velocity, and gate capacitance.

## III. TERMINAL CHARGE MODEL

Proper modeling of the terminal charges is required to account for the dynamic operation of a FET. Under quasi-static conditions, the partitioning of charges at the source ($Q_s$) and the drain ($Q_d$) is accomplished through the Ward-Dutton charge-partitioning scheme [55]; the charge at the gate $Q_g = -(Q_s + Q_d)$; and the derivative of terminal charges with respect to the terminal voltage gives the small-signal capacitances [56]. In a short-channel MOSFET, the carrier transport generally falls somewhere in between the drift-diffusion regime and the ballistic transport regime. The charge model employed in this paper is similar to the VS charge model introduced in [57], in which carrier transport is assumed to be diffusive when $V_{ds}$ approaches zero and ballistic when $V_{ds}$ approaches infinity. The charges in the two extreme cases are computed separately and then combined through a $V_{ds}$-dependent smoothing function. Due to the limited space, the complete derivation of the charge model is detailed in [26, pp. 21-24]. This section focuses on a correction term in the charge model to account for the effect of CNT quantum capacitance ($C_q$).

As illustrated in Fig. 2a, the CNT $C_q$ increases as $V_{gs}$ increases from zero to $V_t$; then reach a maximum; and finally decreases asymptotically to $C_{qinf} \equiv 8q^2/(3a_{cc}\pi E_p)$. The decrease in $C_q$ is because of the rapid drop of CNT DOS after the van Hove singularity [35]. The effect of $C_q$ is not considered in the VS charge model originally developed for silicon MOSFETs. While an analytical model for $C_q$ of CNTs has been developed in [22], the equations are relatively complex, making analytical expressions for $Q_s$ and $Q_d$ hard to obtain. Here, the terminal charge is modeled phenomenologically rather than from first-principles to account for the effect of $C_q$:

$$Q_{ch} = -L_g(Q_{xo} - Q_{xob})$$
$$Q_{xob} = (C_{inv} - C_{invb}) \cdot n_{ss}\phi_t \cdot$$
$$\ln\left(1 + \exp\frac{V_{gs} - [V_{tb} - \alpha \cdot \phi_t \cdot F_f(V_{tb})]}{n_{ss} \cdot \phi_t}\right) \quad (11)$$
$$C_{invb} = C_{ox} \cdot C_{qinf}/(C_{ox} + C_{qinf})$$

where $Q_{ch}$ is the total channel charge, $\phi_t = k_BT/q$ is the thermal voltage, $Q_{xob}$ serves to gradually decrease the absolute value of $Q_{ch}$ around $V_{tb}$, and $V_{tb}$ is a fitting parameter to be determined. $Q_s$ and $Q_d$ are proportional to $Q_{ch}$ as described in [26]. Here we discuss a special case of $V_{ds} = 0$ to demonstrate how the model works. At $V_{ds} = 0$, $Q_s = Q_d = \frac{1}{2}Q_{ch}$, and the small-signal gate capacitance $C_{gg} = -1/L_g \cdot \partial Q_{ch}/\partial V_{gs}$. When $V_{gs} < V_t$, $Q_{xo} \approx 0$, $Q_{xob} \approx 0$, and $Q_{ch} \approx 0$; as $V_{gs}$ increases to $V_t < V_{gs} < V_{tb}$, $|Q_{xob}| << |Q_{xo}|$, so $Q_{ch} \approx -L_gQ_{xo} \approx -L_gC_{inv}(V_{gs}-V_t)$, and $C_{gg}$ approaches the peak value $C_{inv}$; when $V_{gs} >> V_{tb}$, $Q_{xob}$ becomes appreciable and $Q_{ch} \approx -L_g\{C_{inv}\cdot(V_{tb}-V_t)+C_{invb}\cdot(V_{gs}-V_{tb})\}$, and $C_{gg} \approx C_{invb}$, as expected when $V_{gs}$ approaches infinity. The modeled $C_{gg}$ is compared with the numerical simulation [41] in Fig. 9, where $V_{tb} = 0.7E_g/q+0.13$ is determined empirically to achieve the best fitting result. Compared to the case where quantum capacitance is not considered, the $C_{gg}$ including the quantum capacitance is lower and gradually decreases at high $V_{gs}$. The resulting charge model is consistent with the current model because they share the same $V_t$ and $Q_{xo}$.

## IV. CNFET INTRINSIC PERFORMANCE AND CNT DIAMETER

In this section, the impact of CNT diameter on the intrinsic CNFET performance is evaluated based on the model described in Section II and III. Inputs to the VS-CNFET model are: $L_g = 8$ nm, supply voltage $V_{dd} = 0.71$ V, and EOT = 0.51 nm, selected from the "2023" node of the 2013 International Technology Roadmap for Semiconductors (ITRS) projections [58] which predicts the metal-1 pitch will be scaled down to 25.2 nm in 2023 for high performance logic; a GAA structure is assumed; and $R_s = R_Q/2 = h/(2q^2) \approx 3.3$ kΩ per CNT is added to the source and the drain terminals (see Fig. 1) to account for the quantum resistance associated with the interfaces between the 1D CNT channel with the metal S/D contacts (including the lowest band double degeneracy with two spins) [19].

In Fig. 10, the on-state current $I_{on} \equiv I_d(V_{gs} = V_{ds} = V_{dd})$ per CNT and the intrinsic delay $\tau_{int} \equiv L_gC_{inv}V_{dd}/I_{on}$ are plotted against CNT diameter at a fixed off-state current $I_{off} \equiv I_d(V_{gs} = 0, V_{ds} = V_{dd}) = 1$ nA per CNT. As shown in Fig. 10, a 2-nm diameter CNT can deliver 27% higher $I_{on}$ and 21% lower $\tau_{int}$ than a 1-nm diameter CNT. While $\mu \sim d^2$ has been observed experimentally in CNFETs with relatively long channels ($L_g > 4$ μm) [59], here we predict the ratio of $I_{on}(d = 2$ nm$)$ over $I_{on}(d = 1$ nm$)$ to be 1.27, much smaller than $2^2/1 = 4$, because the channel has become nearly ballistic at $L_g = 8$ nm. The increase in $I_{on}$ for large-diameter CNTs is attributed to higher carrier mobility, velocity, and gate capacitance. The advantage of large-diameter CNTs in $\tau_{int}$ is not as prominent as in $I_{on}$ since the gate capacitance is also higher. As will be seen in [32], CNT diameter has greater impacts on the parasitic contact resistance and the tunneling leakage currents in a highly scaled CNFET.

## V. DISCUSSION

The VS carrier velocity is a crucial metric for the transistor technology because it directly determines the magnitude of the drive current as well as the delay of logic devices. A major advantage of the VS model is its capability of extracting $v_{xo}$ directly from the measured data. Normally, the inversion gate capacitance $C_{inv}$ is obtained from the C-V data. Then with $C_{inv}$ as one of the inputs, fitting the VS model to I-V data determines $v_{xo}$ [60]. In other words, both I-V and C-V data are needed in order to reliably extract $v_{xo}$. For emerging devices like CNFETs, however, reliable and reproducible C-V data are often hard to acquire, because of less understanding of CNT-oxide and CNT-metal interfaces and the very small capacitance (aF range) of the 1D channels [61-62]. In this paper, numerical simulation by Sentaurus [52] is used to estimate the $C_{inv}$ as a compromise for the extraction of $v_{xo}$ in Fig. 7. In [27], $v_{xo} = 3 \times 10^7$ cm/s was extracted from a CNFET with $L_g = 9$ nm [2], smaller than the $v_{xo} = 3.8 \times 10^7$ cm/s extracted from the $L_g = 15$ nm CNFET in Fig. 7a. While the contradiction (i.e. $v_{xo}$ of a 9-nm CNFET is smaller than that of a 15-nm CNFET) might be attributed to the differences in gate oxide, fabrication conditions, CNT quality, or the long-range Coulomb interactions described in [63], the unexpected trend highlights the necessity for a larger number of consistent and systematic characterization of devices to extract $v_{xo}$ in CNFETs (e.g. CNFETs built on the same CNT with different gate lengths below 100 nm). These high-quality device data are often not readily available because of the difficulties in device fabrication and the hysteresis and instability of experimental devices.

## VI. CONCLUSION

The intrinsic elements of a compact CNFET model based on the VS approach has been developed in this paper. A VS carrier velocity of $3.8 \times 10^7$ cm/s is extracted from recent experimental CNFET with 15-nm gate length, providing evidence of the superior potential of CNFETs for future transistor technology. The model captures dimensional scaling effects and is used study the impact of CNT diameter on the intrinsic CNFET performance, showing that a 2-nm diameter CNT can deliver 27% higher intrinsic drive current than a 1-nm diameter CNT at $L_g = 8$ nm. The VS-CNFET model has been implemented in Verilog-A and is available online [26]. The model runs smoothly in the SPICE environment (as illustrated in [64]) because all the equations are analytical with no numerical iterations, and the output current is differentiable throughout all regions of operation. A more comprehensive analysis including non-ideal contacts and tunneling leakage is carried out in [32].

## ACKNOWLEDGMENT

The authors would like to thank Prof. Lan Wei (Waterloo) and Prof. Shaloo Rakheja (NYU) for the useful discussion on the VS model; Gage Hills and Prof. Subhasish Mitra at Stanford for testing and using the VS-CNFET model while under development. This work was supported in part through the NCN-NEEDS program, which is funded by the National Science Foundation, contract 1227020-EEC, and by the Semiconductor Research Corporation, and through Systems on Nanoscale Information fabriCs (SONIC), one of the six SRC STARnet Centers, sponsored by MARCO and DARPA, as well as the member companies of the Initiative for Nanoscale Materials and Processes (INMP) at Stanford University.